\documentclass[12pt,preprint]{aastex}

\def\um{\hbox{\,$\mu$m}}
\def\hr{\hbox{$^\mathrm{h}$}}
\def\m{\hbox{$^\mathrm{m}$}}
\def\s{\hbox{$^\mathrm{s}$}}
\def\n2h{\hbox{N$_2$H$^+$}}
\def\kms{\hbox{\,km\,s$^{-1}$}}

\def\msun{\hbox{\,M$_\odot$}}
\def\lsun{\hbox{\,L$_\odot$}}
\def\h13co{\hbox{H$^{13}$CO$^{+}$}}
\def\ks{\hbox{$K_s$}}
\def\cm3{\hbox{\,cm$^{-3}$}}
\def\cmm2{\hbox{\,cm$^{-2}$}}

\def\mjy{\hbox{\,mJy\,beam$^{-1}$}}
\def\nh{\hbox{N$_\mathrm{H}$}}
\def\av{\hbox{A$_\mathrm{V}$}}

\slugcomment{}

\shorttitle{A dense micro-cluster of Class\,0 protostars in NGC\,2264\,D-MM1}
\shortauthors{Teixeira, Zapata, \& Lada}

\begin{document}

\title{A dense micro-cluster of Class\,0 protostars in NGC\,2264\,D-MM1}

\author{Paula S. Teixeira\altaffilmark{1,2,3}, Luis A. Zapata\altaffilmark{4} and Charles J. Lada\altaffilmark{1}}
\altaffiltext{1}{Harvard-Smithsonian Center for Astrophysics, 60 Garden Street, Cambridge, MA 02138; pteixeira@cfa.harvard.edu, clada@cfa.harvard.edu}
\altaffiltext{2}{Departamento de F\'{\i}sica da Faculdade de Ci\^encias da \hbox{Universidade} 
de Lisboa, Ed. C8, Campo Grande, 1749-016, \hbox{Lisboa},Portugal}
\altaffiltext{3}{Laborat\'orio Associado Instituto D. Luiz - SIM, 
Universidade de Lisboa, Campo Grande, 1749-016, \hbox{Lisboa}, Portugal}
\altaffiltext{4}{Max-Planck-Institut f\"ur Radioastronomie, Auf dem Hügel 69, 53121 Bonn, Germany; lzapata@mpifr-bonn.mpg.de}

\begin{abstract}
We present sensitive and high angular resolution ($\sim$ 1\arcsec) 1.3\,mm continuum observations of the dusty core D-MM1 in the Spokes cluster in NGC\,2264 using the Submillimeter Array. A dense micro-cluster of seven Class\,0 sources was detected in a 20\arcsec $\times$ 20\arcsec\ region with masses between 0.4 to 1.2\msun\ and deconvolved sizes of about 600\,AU. We interpret the 1.3\,mm emission as arising from the envelopes of the Class\,0 protostellar sources. The mean separation of the 11 known sources (SMA Class\,0 and previously known infrared sources) within D-MM1 is considerably smaller than the characteristic spacing between sources in the larger Spokes cluster and is consistent with hierarchical thermal fragmentation of the dense molecular gas in this region.

\end{abstract}


\section{Introduction}
\label{sec:intro}

NGC\,2264 is a very well studied hierarchically structured young embedded cluster \citep{lada03} 
associated with a giant molecular cloud in the Monoceros OB1 complex, 800\,pc \citep[$\pm$50\,pc;][]{sung97,peretto06} away 
from the Sun. There is abundant evidence of ongoing star forming activity in this cluster, 
such as molecular outflows \citep{margulis88,wolf-chase03} and Herbig-Haro objects \citep{reipurth04}. 
\citet{margulis89} identified luminous far infrared sources with the InfraRed Astronomical Satellite (IRAS), 
many of them being Class\,0 and Class\,I objects. Submillimeter observations of these 
sources by \citet{williams02}, \citet{wolf-chase03}, and \citet{peretto06} identified 
several dense clumps embedded in dusty filamentary fingers of molecular material that radiated out from IRAS\,12. 
More recent observations using the \emph{Spitzer} Space Telescope revealed that many of these clumps 
contained bright 24\um\ Class\,0/I sources and the clustering was named the Spokes cluster due 
to its geometrical configuration \citep{teixeira06}. Using nearest neighbor analysis, \citet{teixeira06} 
found that these protostars have a characteristic spacing that corresponds to the Jeans length 
and are very likely tracing the primordial substructure of the cluster. 
They interpret the regular spacing of the protostars as a fossil signature of thermal fragmentation. 
The most massive and densest of these submillimeter cores, D-MM1 (also known as IRAS\,12\,S1) has 
a mean density of $\sim$10$^6$\cm3\ \citep{wolf-chase03,peretto06}. HCO$^{+}$ observations of D-MM1 
by \citet{williams02} shows a deep red-shifted absorption indicating that it is collapsing at a speed 
that \citet{peretto06} measured to be $\sim$0.1\kms. \citet{wolf-chase03} identified D-MM1 as a Class\,0 object harbouring one or more protostars. Recent near-infrared imaging using the PANIC camera 
at the Magellan telescope, as well as \emph{Spitzer} IRAC imaging resolved D-MM1 into 11 \ks-band and 8 8\um\ sources  \citep{young06} within the field we observed with the Submillimeter Array (SMA). These high angular resolution observations resolved D-MM1 into a dense micro-cluster of 7 millimeter compact sources associated 
with young Class\,0 objects. We describe our observations in 
Section \S\,2 and present our results in Section \S\,3. Finally, we discuss our findings on the 
characterization of these new millimeter sources in Section \S\,4.

\section{Observations}
\label{sec:obs}

The observations were made with the Submillimeter Array (SMA) during 2005 December 25 in its ``compact'' 
configuration, and 2006 February 5 and 10 in its ``extended'' configuration, with eight antennas in both 
configurations.
The frequency was centered at 230.538\,GHz in the lower sideband, while the upper 
sideband was centered at 220.538\,GHz. The primary-beam size (half-power beamwidth) of the 6\,m diameter 
antennas at 230\,GHz is 54\arcsec. 
The phase reference centers of the field is \hbox{($\alpha,\delta$)(J2000)=(06\hr41\m06.45\s, 
+09\degr33\arcmin47.91\arcsec)}.
The zenith opacity, ($\tau_{230\,\mathrm{GHz}}$), measured with the National Radio Astronomy Observatory  (NRAO)
tipping radiometer located at the Caltech Submillimeter Observatory, varied between 0.03 to and 0.04 
in the extended configuration observation time, and in the compact configuration observation time was 
very stable of 0.2.
The phase and amplitude calibrators were the quasars 0739+016 and 0530+135 with measured flux densities of 
2.82 $\pm$ 0.1\,Jy and 0.84 $\pm$ 0.1\,Jy, respectively. The uncertainty in the flux scale is estimated 
to be 20\%, based on the SMA monitoring of quasars. Observations of Uranus provided the absolute scale for 
the flux density calibration. Further technical descriptions of the SMA are found in \citet{ho04}.
The data were calibrated using the Interactive Data Language superset MIR, based on the MMA software 
developed for the Owens Valley Radio Observatory \citep{scoville93} which was adapted for the SMA\footnote{\url{http://www.cfa.harvard.edu/$\sim$cqi/mircook.html}}. 
The calibrated data were imaged and analyzed in the standard manner using the Multichannel Image Reconstruction 
Image Analysis and Display (MIRIAD) package. We weighted the \emph {(u,v)} \rm data using the ROBUST parameter 
of INVERT MIRIAD task set to 2, optimizing for a maximum sensitivity in the continuum image. This option is recommended 
to achieve the largest signal-to-noise ratio possible, although some angular resolution is sacrificed.
The synthesized beam had dimensions of $1\farcs4 \times 1\farcs3$ with a P.A. = 10.4\degr. 
Finally, the resulting continuum image rms noise level ($\sigma$) was 3.5\mjy.

\section{Results: new millimeter sources}
\label{sec:res}

\begin{figure*}
\centering
\includegraphics[width=0.75\textwidth]{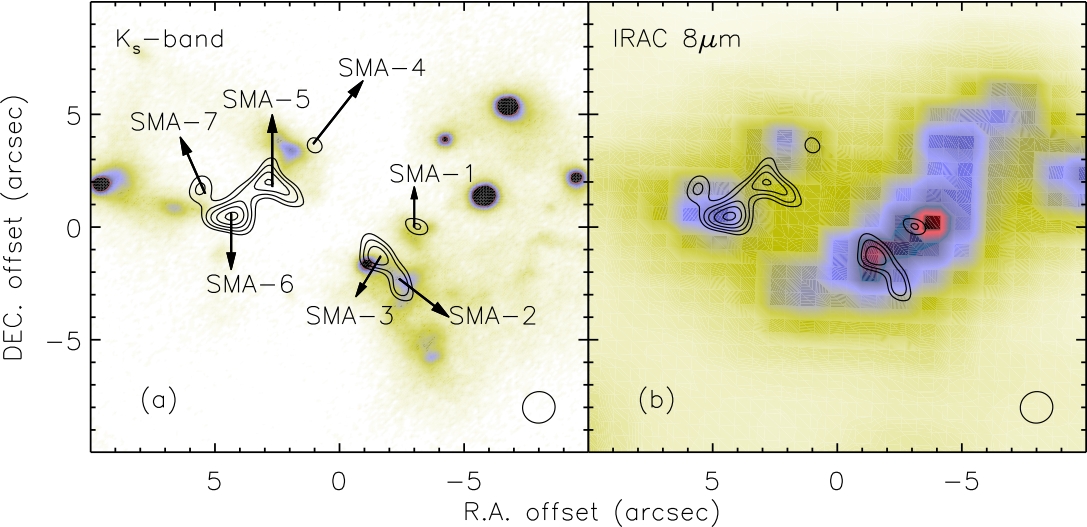}
\caption{K$_\mathrm{s}$-band (a) and 
 \emph{Spitzer} IRAC 8\um\ (b) images \citep{young06} of D-MM1. The images are centered on \hbox{$(\alpha,\delta)$(J2000)=(06\hr41\m05.8\s, +09\degr34\arcmin08.8\arcsec)}. 
The SMA continuum data is overlayed in black contours ranging from 5 to 9 sigma (3.5\mjy) in both panels. 
The ellipse in the lower right corner shows the FWHP synthesized beam.}
\label{fig:sma1}
\end{figure*}

We detected seven compact sources whose peak fluxes exceeded 5 $\sigma$ within the core D-MM1. Figure \ref{fig:sma1} shows a panel of near-infrared (NIR) and mid-infrared (MIR) images of the core with the SMA sources overplotted (with contours). 
The \ks-band image (0.125\arcsec pixel$^{-1}$), depicts emission in the vicinity of the SMA sources, however this NIR emission is generally not pointlike (except for the source associated with SMA-3) and may correspond to scattered light from a central protostar. 
The \emph{Spitzer} Infrared Array Camera (IRAC) image \citep{teixeira06} (1.2\arcsec pixel$^{-1}$), shows that \hbox{SMA-1} and \hbox{SMA-3} have MIR counterparts. The sources \hbox{SMA-2}, \hbox{SMA-6}, and \hbox{SMA-7} are also associated with diffuse MIR emission.
The point-like sources detected in the \ks-band and the 8\um\ images, that were not detected by our SMA observations, had been previously identified as Class\,I sources by \citet{young06} (sources \# 30, 33, 35, 44, 45, 49).
In Table \ref{tab:prop} we present the physical parameters of these new objects. 
To measure the fluxes for the SMA sources, we used the NRAO Astronomical 
Image Processing System (AIPS) software. Flux densities, as well as source positions and sizes,  
were obtained using the AIPS IMFIT procedure, where each source was modeled with two-dimensional 
elliptical Gaussians. The errors cited for the flux densities correspond to statistical uncertainties 
and are dominated by the calibrational uncertainty that is $\sim$\,20\% as previously mentioned.

To determine the mass we assume the emission arises from an optically thin, unresolved, source that can be 
characterized by a single temperature \citep[e.g.][]{mundy95,bally98}. 
The mass is calculated by \hbox{$M=F_\nu d^2 /(B_\nu(T_d) \kappa_\nu)$}
where $\nu$ is the frequency, $T_\mathrm{d}$ is the dust temperature, $F_\nu$ is the integrated flux density of the source, 
$B_\nu$ is the Planck function,
and finally, $\kappa_\nu$ is 
the dust mass opacity. 
We derive the latter value from \citet[][$\kappa_\nu=0.1 (\nu/1200\,\mathrm{GHz})^\beta$\,cm$^2$\,g$^{-1}$]
{beckwith90} as being \hbox{$\kappa_\nu=(0.0192)^\beta$\,cm$^2$\,g$^{-1}$}. The value of $\kappa_\nu$ assumes a gas-to-dust ratio 
of 100 \citep{hildebrand83}, which may not be the most adequate to use for protostellar sources since dust settling to the 
mid-plane of the disk and erosion of the circumstellar envelope by photo-dissociation may decrease the gas-to-dust ratio 
\citep{williams05,throop05}. Additionally, the value of the emissivity index, $\beta$, is uncertain \citep{beckwith90}. 
The sizes of the sources suggest we are observing emission from compact envelopes, hence we adopt $\beta=1.5$ to calculate the mass of the sources \citep{wolf-chase03,peretto07}. We use 23\,K for the mass determination, as this was temperature measured by \citet{wolf-chase03} for D-MM1.
Table \ref{tab:prop} shows the masses we determined, as well as the average number density, $\bar{n}$, given by 
\hbox{$\bar{n}=3M/(4 \pi \mu m_\mathrm{H} R_{geom}^3)$}, 
where $\mu$ is the mean molecular weight, 2.34, $m_\mathrm{H}$ is the atomic H mass, and $R_{geom}$ is the geometric mean 
radius. Due to the uncertainties referred to above the values of the derived masses are good within a factor of two.
The values of $\bar{n}$ should be interpreted only as an estimate since the error associated with the size of 
the source is uncertain and reflects the particular ($u,v$) coverage of the data. 

\section{Discussion}
\label{sec:diss}

\subsection{On the nature of the new micro-cluster}
\label{subsec:efficiency}

Following the number counts of extragalactic objects at millimeter wavelengths from \citet{maloney05},
we estimate that the expected number of 1.3\,mm background sources, above a flux density of S=20\,mJy, is 
\hbox{$\langle$ N $\rangle$ $\sim$ 5.76$^{+0.30}_{-9.1}$ deg$^{-2}$}.
The probability of finding a source with flux density equal or larger than 20\,mJy 
(see Table \ref{tab:prop}) in a 20\arcsec $\times$ 20\arcsec\ region is thus quite low,  
$\sim$ 10$^{-4}$. We therefore conclude that all the observed SMA sources are associated with NGC\,2264\,D-MM1 core.
We find a very similar value using the number counts for submillimeter galaxies reported by \citet{laurent05}.
Out of the seven millimeter continuum compact sources, two are associated with pointlike emission and three are associated with diffuse emission.
 (see Fig. \ref{fig:sma1} and Table \ref{tab:prop}). 
Sources SMA-4 and SMA-5 do not appear to have any NIR or MIR emission, although there is NIR and MIR diffuse emission located between these sources (which could be scattered light from one or from both sources). 
The deconvolved sizes ($\sim$ 600 AU) of these objects, as well as the high densities ($\sim$10$^8$) suggest the 1.3\,mm emission is arising from compact dense regions such as circumstellar envelopes. The derived masses (0.4 to 1.2\msun) are comparable to those of low mass stellar objects indicating that a substantial amount of the total protostellar material is in the circumstellar envelopes, from which we conclude that the SMA sources are very likely protostars in the Class\,0 evolutionary phase. 
We cannot advance more in our interpretation of the individual sources due to a lack of complementary high resolution observations at infrared and (sub)millimeter wavelengths of this region (namely at 850\um). We would require such observations to build individual spectral energy distributions and model these to determine  protostellar envelope and disk masses. 
We are however able to discuss the fragmentation of D-MM1 and the formation of these sources. There are 11 sources within D-MM1 \citep[excluding one foreground star][]{young06}, 7 of which are the SMA sources we are reporting in this Letter. These sources are more densely packed than those in the Spokes cluster, which have a characteristic spacing similar to the Jeans length \citep{teixeira06}.
Our interferometric observations recover $\approx$90\% of the flux of D-MM1 measured by \citet{peretto06} using their background subtracted maps. This implies that the single dish observations correspond to the convolution of the individual fluxes emitted by the SMA sources \citep[as suggested by][]{wolf-chase03}. If we assume that each protostar has 0.5\msun, then a rough estimate of the pre-fragmentation density of D-MM1 can be calculated by adding up the measured masses of the envelopes and the adopted protostellar masses, which gives us 10.9\msun, and distributing this total mass within a region of 7870\,AU radius \citep[][]{peretto06}. The density we estimate in this manner is 8$\times$10$^5$\cm3. 
The Jeans length, $\lambda_\mathrm{J}=\sqrt{(\pi k_\mathrm{B} T)/(G\bar{n})}/(\mu m_\mathrm{H})$ ($k_\mathrm{B}$ is the Boltzmann constant and $G$ is the gravitation constant) we obtain for D-MM1 is 5.9\arcsec\ (0.023\,pc or 4740\,AU). Here we use a temperature T of 10\,K since it is logical to assume that the pre-fragmented starless core would be colder than the present temperature of D-MM1, 23\,K. The corresponding Jeans length is very similar to the mean distance between the 11 sources, 6.9\arcsec\ (0.026\,pc or 5520\,AU). However, the mean nearest neighbor se\-pa\-ra\-tion between the sources is 2.3\arcsec\ (0.009\,pc or 1840\,AU), so it is not entirely clear if the fragmentation of D-MM1 was purely thermal. It is interesting to compare our results for D-MM1 with the northern subgroup of the Serpens cluster, which consists of a rich clustering of Class\,0 and Class\,I sources. Placing these Serpens sources at the distance of NGC\,2264 and scaling the 1.4\,mm fluxes measured by \citet{hogerheijde99} accordingly gives us fluxes comparable to what we are measuring for our SMA sources.
\citet{winston07} found that the average separation of the Serpens protostars is about 5000\,AU with some of the sources being as close as 2000\,AU. They also calculate the Jeans length for the region to be 0.024\,pc, which is similar (neglecting projection effects) to what we find. If this is a characteristic scale then this could be an indication that thermal physics is the underlying engine that regulates star formation, in at least some highly clustered environments.
Given that the nearest neighbor separations of the sources in D-MM1 are smaller than typical envelope sizes (5000 to 10000\,AU), these sources may be competing between themselves for the accretion of the surrounding material of the core. Therefore, we cannot rule out the possibility that some kind of competitive accretion \citep{bonnell01a, peretto06} may be occurring in D-MM1. Additional data is required, such as the measurement of the velocity dispersion of the sources, before we are able to draw any conclusions regarding this issue.
Our finding holds within it an interesting implication, specifically, that there are two scales of fragmentation in the Spokes cluster: one that formed D-MM1 and other bright members of the cluster \citep{teixeira06} and a second associated with the fragmentation of the D-MM1 core itself. These two fragmentation length scales are correlated with the mean density in those regions, as would be expected if fragmentation was dominated by thermal physics.
The Jeans mass for D-MM1, $M_\mathrm{J}=17T^{3/2}\bar{n}^{-1/2}$\msun, is 0.6\msun.  Since D-MM1 would have had 10.9\msun\ in its pre-fragmented state, we expect the core to have fragmented into about 17 sources if thermal pressure was the dominant support against gravitational collapse. This value is within a factor of two of the number of sources in D-MM1, 11.
A more detailed analysis is obviously required in determining the underlying stellar masses. It is however interesting to note that if the Class\,0/I sources in D-MM1 have in fact M\,$\sim$\,0.5\msun, then their bolometric luminosities could be similar to those of IRAS\,05173-0555, RNO\,43\,MM, CB\,230, IRAS\,18148-0440, or RNO\,15\,FIR, i.e., 7-10\lsun\ \citep{froebrich05}, which would yield a combined luminosity of 77-110\lsun. The bolometric luminosity of D-MM1 is 107.5\lsun\ \citep{wolf-chase03}.

\subsection{X-ray emission associated with SMA-1}
\label{subsec:x-ray}

\citet{flaccomio06} used the Advanced CCD Imaging Spectrometer \hbox{(ACIS)} onboard the Chandra X-ray Observatory (CXO) 
to obtain a 97\,ks long exposure of NGC\,2264. We searched their X-ray catalog and found that source \# 237 
(\hbox{CXO ANC J064105.5+093407}: \hbox{($\alpha, \delta$)(J2000)=(06\hr41\m05.55\s, +09\degr34\arcmin07.68\arcsec)} 
is located approximately 0.6\arcsec\ away from SMA-1. We note that the SMA has subarcsecond pointing accuracy and the 
CXO has a positional accuracy of typically 0.6\arcsec. To our present knowledge there are no other infrared or millimeter 
sources positioned closer to \hbox{CXO ANC J064105.5+093407} so it could potentially be associated with SMA-1. 
The ACIS data has been recently re-analyzed and made publicly available through AN archive of CHandra Observations of 
Regions of Star formation (ANCHORS) \citep[http://cxc.harvard.edu/ANCHORS/;][]{wolk06}. ANCHORS is a web based 
archive of all the point sources observed during Chandra observations of regions of star formation and their fitted spectra.
Although the number of photons detected (9) was very low it is nontheless illustrative to mention the spectral fit obtained to help gain some insight into the nature of the source. 
The fitted spectra of \hbox{CXO ANC J064105.5+093407} corres\-ponds to an absorption hydrogen column density 
of \nh=16.58$\pm$5.21$\times$10$^{22}$\cmm2, kT=0.90$\pm$0.13\,keV (i.e., a plasma temperature of 10.4$\pm$1.5\,MK), 
and a 0.3 solar metal abundance.
We derive an equivalent visual extinction \av\ of  103$\pm$32 magnitudes using the gas-to-dust ratio 
$N_\mathrm{H}(\mathrm{cm}^{-2})=1.6\times10^{21}A_\mathrm{V}$(mag) \citep{vuong03}. Dust extinction maps of NGC\,2264 using the near-infrared color excess 
method \citep[e.g.][and references therein]{teixeira05} show that the region of the Spokes cluster is highly extincted, 
with \av\ $>$ 40\, magnitudes, unpublished data. So the measured \av\ of \hbox{CXO ANC J064105.5+093407} is consistent
 with it being either an Active Galaxy Nucleus (AGN) observed through the dense molecular cloud or a deeply 
embedded young stellar object. 
The fitted spectra has very little emission in the soft X-ray energy regime (0.5-2\,keV) and is mostly hard 
X-ray (2-10\,keV), peaking at 3\,keV. AGN spectra are characterized by hard X-ray emission \citep[e.g.][]{mushotzky00},
 however, the plasma temperature derived from the fit is below that which is expected for AGNs (T$_\mathrm{plasma AGN} \ga 10^9$\,K)
 and very similar to what  is observed towards protostellar objects in the Coronet cluster \citet{forbrich06}. 
Therefore \hbox{CXO ANC J064105.5+093407} is associated with 
a protostellar object, very likely SMA-1. \citet{winston07} also reports Class\,0/I sources with X-ray emission.
Unfortunately the CXO ACIS-I camera was pointed and orientated such that D-MM1 fell precisely in the gap between the chips of the array: this may be the reason why no other X-ray source is detected in D-MM1.

\section{Summary}
\label{sec:summ}

We discovered a dense micro-cluster of Class\,0 sources within D-MM1, with envelope masses ranging from 0.4 to 1.2\msun\ and average radius of 600\,AU.
Five of the sources are associated with NIR/MIR emission, and one of the SMA sources, SMA-1, is associated with X-ray emission. The separations of the sources indicate that the fragmentation length scale of this core is significantly smaller than that of the Spokes cluster and comparable to the Jeans length in D-MM1. This is consistent with hierarchical thermal fragementation: with a primary fragmentation along the lower density (10$^4$\cm3) filaments of the Spokes cluster, and a secondary fragmention of the higher density core D-MM1 ($\sim$10$^6$\cm3).\\

\acknowledgments

The SMA is a joint project between the SAO and the Academia Sinica Institute of Astronomy and Astrophysics and is funded by the Smithsonian Institution and the Academia Sinica.
P. S. T. acknowledges the scholarship SFRH/BD/13984/2003 from the Funda\c{c}\~ao para a Ci\^encia e Tecnologia (Portugal).

{\it Facilities:} \facility{SMA}.


\clearpage

\clearpage

\begin{deluxetable}{cccccccccccc}
\tablecolumns{12}
\tablewidth{0pt}
\tabletypesize{\scriptsize}
\tablecaption{Properties of the SMA sources\label{tab:prop}}
\tablehead{
\colhead {} & \multicolumn{7}{c}{Measured} & \colhead {} & \multicolumn{3}{c}{Calculated}\\
\cline{2-8} \cline{10-12} \\
\colhead{ID.} & \colhead{R.A.\tablenotemark{(a)}} & \colhead{DEC.\tablenotemark{(a)}} & \colhead{Size\tablenotemark{(b)}}   & \colhead{P.A.} & \colhead{F$_\mathrm{230\,GHz\ peak}$} & \colhead{F$_\mathrm{230\,GHz\ int.}$}  & \colhead{NIR} & & \colhead{R$_\mathrm{geom}$} & \colhead{Mass} & \colhead{$\bar{n}$}  \\
\colhead{} & \colhead{06\hr41\m} & \colhead{09\degr34\arcmin}  & \colhead{(\arcsec)} & \colhead{(\degr)} & \colhead{(mJy\,beam$^{-1}$)}  & \colhead{(mJy)}  & \colhead{/MIR\tablenotemark{(c)}}  & & \colhead{(AU)} &  \colhead{(\msun)} & \colhead{($10^7$\cm3)} 
}
\startdata
1 & 05.60\s  & 08.0\arcsec & 1.8($\pm$0.6)$\times$0.7\tablenotemark{d} & 70$\pm$13 & 21$\pm$6 & 29$\pm$14  & Nd, Mp   & & 448  & 0.4$\pm$0.2 & 14.7 \\
2 & 05.64\s  & 05.7\arcsec & 3.1($\pm$0.7)$\times$0.7\tablenotemark{d} & 23$\pm$6  & 24$\pm$6 & 56$\pm$19  & Nd, Md   & & 592  & 0.7$\pm$0.2 & 12.2 \\
3 & 05.69\s  & 06.7\arcsec & 2.6($\pm$0.6)$\times$0.9($\pm$0.6)	       & 42$\pm$13 & 27$\pm$6 & 69$\pm$21  & Np, Mp   & & 608  & 0.9$\pm$0.3 & 13.8 \\
4 & 05.87\s  & 11.7\arcsec & 1.8($\pm$1.0)$\times$1.2($\pm$1.2)	       & 96$\pm$45 & 19$\pm$6 & 41$\pm$19  & none     & & 584  & 0.5$\pm$0.2 & 9.2 \\
5 & 05.98\s  & 09.8\arcsec & 2.6($\pm$0.6)$\times$1.6($\pm$0.5)	       & 59$\pm$32 & 27$\pm$6 & 90$\pm$25  & none     & & 816  & 1.1$\pm$0.3 & 7.5 \\
6 & 06.09\s  & 08.6\arcsec & 2.5($\pm$0.5)$\times$1.5($\pm$0.4)	       & 87$\pm$18 & 30$\pm$6 & 94$\pm$24  & Md       & & 776  & 1.2$\pm$0.3 & 9.0 \\
7 & 06.16\s  & 09.5\arcsec & 2.1($\pm$0.6)$\times$1.2($\pm$0.6)	       & 10$\pm$57 & 21$\pm$6 & 51$\pm$20  & Md       & & 632  & 0.6$\pm$0.2 & 9.1 \\
\enddata
\tablenotetext{(a)}{J2000.}
\tablenotetext{(b)}{deconvolved FWHM size derived from fitting a 2D elliptical Gaussian to the continuum SMA sources.}
\tablenotetext{(c)}{Nd(Md): associated with diffuse \ks-band(8\um) emission; Np(Mp): associated with pointlike \ks-band(8\um) emission.}
\tablenotetext{(d)}{the minor axis is unresolved and we use an upper limit of beamsize/2.}
\end{deluxetable}

\end{document}